# Improvising Age Verification Technologies in Canada: Technical, Regulatory and Social Dynamics

Azfar Adib, *Senior Member, IEEE*
*Dept. of Electrical and Computer Eng.*
*Concordia University*
Montreal, Canada
azfar.adib.eee@gmail.com

Wei-Ping Zhu, *Senior Member, IEEE*
*Dept. of Electrical and Computer Eng.*
*Concordia University*
Montreal, Canada
weiping@ece.concordia.ca

M. Omair Ahmad, *Fellow, IEEE*
*Dept. of Electrical and Computer Eng.*
*Concordia University*
Montreal, Canada
omair@ece.concordia.ca

*Abstract*— Age verification, which is a mandatory legal requirement for delivering certain age-appropriate services or products, has recently been emphasized around the globe to ensure online safety for children. The rapid advancement of artificial intelligence has facilitated the recent development of some cutting-edge age-verification technologies, particularly using biometrics. However, successful deployment and mass acceptance of these technologies are significantly dependent on the corresponding socio-economic and regulatory context. This paper reviews such key dynamics for improvising age-verification technologies in Canada. It is particularly essential for such technologies to be inclusive, transparent, adaptable, privacy-preserving, and secure. Effective collaboration between academia, government, and industry entities can help to meet the growing demands for age-verification services in Canada while maintaining a user-centric approach.

*Keywords*— *Age verification, biometrics, accessibility, inclusiveness, PCTF, PAS 1296:2018, IEEE STD 2089-2021.*

## I. BACKGROUND

### A. Need for Age Verification

Age verification indicates validating users' age levels before delivering certain age-appropriate services or products. In a general context, an age check is required for commodities available for primarily three categories of people- minors, adults, and seniors. For instance, senior citizens (65 years or above) in many countries enjoy certain benefits in public transit, medical facilities, and recreational parks. To avail of those, they need to show proof of their ages. The other context is segregating adults and minors, which is the biggest used case of age verification. This is required for products and services available only for adults, such as tobacco, gambling, cannabis, etc.

For such traditional age-restricted products (gambling, tobacco, alcohol), age restriction-related laws and regulations have existed for a long time in different countries. For example, several members of the OECD countries (e.g., Germany, France, Spain, Italy, and Australia) have restrictions prohibiting access for under-aged people etc. In countries like the United States, Canada, and India, different laws exist at state/province levels regarding such restrictions. In places like China and Russia, uniform law exists for all parts of the country [1].

Rapid digitalization and widespread internet connectivity around the world have sparked a massive increase in age-dependent services in the online domain. This first includes the online purchase of traditional age-dependent services. Secondly, it covers a huge range of online content which can be accessed only after a certain age. For instance, all social media platforms have a minimum age requirement. This age limit is usually set by the corresponding platform, being independent of state-driven regulations. Online gaming is another significant case of age verification.

TABLE I.  MINIMUM AGE REQUIREMENTS IN SOME POPULAR ONLINE PLATFORMS [2]

| Minimum Age Limit (Years) | Online Platforms |
| --- | --- |
| 13 | Facebook, Instagram, Twitter, Pinterest, Google+, Snapchat, Discord, Tiktok |
| 14 | LinkedIn |
| 16 | WhatsApp |
| 18 | YouTube, Flickr, WeChat |

Certain adult contents require strict age assurance so that minors do not get exposed to undesirable content. This has been a widely discussed issue recently around the world, fueled by the exponential increase of harmful occurrences impacting minors (especially women). Different countries are passing regulations to tackle this challenge. In 2017, the United Kingdom passed the 'Digital Economy Act 2017' and became the first country in the world to enforce a legal mandate on the Internet age verification system [3]. The UK is currently reviewing the "Online Safety Bill," which includes an updated set of laws to protect children online [4]. The Senate of Canada recently passed Bill S-210, which proposes to restrict young persons' online access to sexually explicit material by enforcing age verification on relevant websites [5]. France introduced a similar bill in 2020 to implement a nationwide age verification system for pornography websites [6]. Around 18 states in the US have already introduced or passed age-verification-related bills, while it has become a prioritized issue for the federal government as well [7].

### B. Methods of Age Verification

From a general perspective, age verification schemes can be divided into below categories [2][7]:

1. Self-check: Users are asked to declare their ages while agreeing to the terms of usage of the service. Popular social

media platforms (like Facebook, Viber, SnapChat) use such a policy. The effectiveness of such a scheme is questionable, as it does not verify the declared age of the users.

2. Document-based verification: Users share digital versions of authorized identification documents (like driving licenses, and health cards) to prove their ages. Following the trend of the physical world, this has been the most common method so far for online age verification also.

3. Biometric-based verification: Users' ages are determined by their biometric features, such as facial features, voice, height, gate, fingerprint etc.

4. Capacity testing: Capacity testing allows a service to estimate a user's age based on an assessment of their aptitude or capacity. For example, a child may be asked to complete a language test, solve a puzzle or undertake a task that gives an indication of their age or age range.

5. Cross-account or third-party verification: Users can verify their ages through an existing verified account in another domain or through a third party (like- a credit card, or an age-certificate), through which their ages get cross-checked.

6. Profiling: Profiling refers to the process of analyzing online behavior of users to predict their ages. Data used for profiling consist of information users choose to share about themselves and information that is automatically collected from their engagement with services. For example, how long they spend on a webpage, where their cursor hovers, the times of day they access services and their interests, location, and friends.

7. Authorized confirmation: In this approach, an adult account holder is either asked to confirm the age of the child user, or they may be asked to set up a special account for the child. In both scenarios, the adult takes responsibility for verifying the age of the child.

Usually, document-based, and cross-account/third-party verification schemes are commonly used methods of age verification. However, these methods possess some risks regarding document forgery, along with significant privacy concerns. Since the start of the COVID-19 pandemic, the exponential increase in online transactions has fueled the need for a secure and reliable mechanism of age verification. At the same time, increased awareness is visible among users regarding their data privacy and security. For instance, in a survey carried out by Interac in January 2023 among adult Canadians, 76% respondents expressed concerns about protecting their online privacy. They are particularly worried about sharing a digitalized copy of their physical ID. These concerns re-emphasize the need for schemes that can protect the privacy of users while ensuring accurate age verification.

*C. Age Verification Using Biometrics*

The rapid progress of machine learning in recent years has facilitated novel innovation in age verification using biometrics. Facial features are so far considered as a promising avenue for age recognition. For instance, Yoti, the leading age verification service provider in the world, has developed facial age estimation based on neural networks. For 6-12-year-olds, they obtained a True Positive Rate (TPR) of 98.91% for getting correctly estimated as under 13 and a mean absolute error (MAE) of 1.36 years. For 13-17-year-olds, TPR was 98.65% for being correctly as estimated under 23, and MAE was 1.52 years. This technology is being deployed in various places. For instance, since March 2023, Instagram has started piloting facial age estimation technology in Canada which they have already tested in the US, UK, Brazil and India. In their scheme, the users can either upload a photo identification or they can record a video selfie [9]. Speech-based age recognition has also been explored by some researchers. Safavi et al. analyzed speech recognition for age-group identification for children aged 5-16, achieving an accuracy of 85.8%. However, speech may not be a reliable feature for age recognition for children aged 11-13 due to variations during the onset of puberty. Also, it has the risk of forgery by playing voice recordings [10]. Falohan et al. used fingerprint characteristics (such as the Ridge Thickness, Valley Thickness Ratio, and the Ridge Count) for age group recognition. While that approach provides high accuracy (82.14%) and is hard to circumvent, it requires a fingerprint reader [11]. It also does not guarantee the anonymity of the user. Yaman et al. explored ear features extracted from anthropometric landmarks of the ear (distance measurements and area calculations) for age classification. However, ear features are not a very decisive characteristic in a person's growth and have lower accuracy compared to the other techniques [12]. Iris images were used by Erbilek et al. to distinguish users between different age groups using a combination of a small number of geometric features. However, it requires the use of a near-infrared camera which may not always be available on a mobile device. This also may not ensure anonymity [13]. Electrocardiogram (ECG), which has emerged during the last few years as promising biometrics, has also been explored for age verification. Using a scheme consisting of discrete wavelet transformation and one dimensional convolutional neural network, Adib et al. achieved up-to 99% classification accuracy in segregating adults and minors using ECG [14].

## II. ONLINE AGE VERIFICATION: CANADIAN CONTEXT

*A. Usage Context*

Like the rest of the world, internet usage in Canada has been increasing rapidly during the last two decades. In 2020, 94% of Canadians had household internet access through a fixed broadband connection. 80% of Canadians reported having a mobile data plan. Around 92% of the internet users in Canada use social media [15]. Facebook is the most popular social media platform here. TikTok is considered the fastest growing one, despite recent bans imposed by federal and provincial governments across Canada regarding its use in government devices. The top five social media platforms in Canada are Facebook, Instagram, TikTok, Twitter, and Pinterest. The top five messenger Apps are Facebook Messenger, WhatsApp, iMessage, Snapchat, and Discord. The age group between 35 to 39 years encompasses the highest proportion of social media users in Canada, as depicted in table II [16].

TABLE II. AGE-WISE COMPOSITION OF SOCIAL MEDIA USERS IN CANADA

| Age Group | Percentage |
|---|---|
| 15 to 19 | 9.9 |
| 20 to 24 | 11.0 |
| 25 to 29 | 12.1 |
| 30 to 34 | 11.8 |
| 35 to 49 | 30.3 |
| 50 to 64 | 24.8 |

Any specific statistics regarding online purchases of other age-restricted products (e.g., tobacco, alcohol, cannabis) were not found. However, all these products have a huge volume of yearly sales. During 2021-2022, sales of alcoholic beverages in Canada were 26.1 Bn USD. For cannabis, this figure was 4 Bn USD. In the same period, Canadians spent 11.3 Bn USD on gambling [17] [18].

Any official statistics regarding adult site usage in Canada are also not available. As reported by Mindgeek, a Canada-based company that runs some known adult content sites (like "Pornhub"), Canada ranks 8[th] in the world in terms of its visitor traffic on that site. The age group between 25 to 34 years encompasses the highest proportion of these visitors, as shown in table III [18].

TABLE III. AGE-WISE COMPOSITION OF CANADIAN USERS IN A KNOWN ADULT SITE

| Age Group | Percentage |
|---|---|
| 18 to 24 | 22 |
| 25 to 34 | 27 |
| 35 to 44 | 20 |
| 45 to 54 | 14 |
| 55 to 64 | 10 |
| 65+ | 7 |

*B. Legal Context*

Being a federation of 10 provinces and 3 territories, laws-regulations in Canada are divided among federal, provincial (and also municipal) authorities. For federal laws, which apply to every Canadian regardless of the province or territory, the age of majority is 18. This indicates the age when a Canadian citizen can vote and contest elections, and they receive access to certain rights-services. Another aspect of federal laws that are coming into effect is the age-restricted access to adult content. It is Bill S-210, termed as "An Act to restrict young persons' online access to sexually explicit material," which got passed in the Senate of Canada in April 2023 and is currently on review in the House of Commons. It has specified 18 years as the minimum age requirement for accessing such content [5].

For traditional age-restricted products, all provinces and territories exercise their own regulations. A summary of that is given in

TABLE IV. PROVINCIAL REGULATIONS IN CANADA REGARDING MINIMUM AGE LIMIT (YEARS) [20]

| Province | Alcohol | Tobacco | Cannabis |
|---|---|---|---|
| Alberta | 18 | 18 | 18 |
| British Columbia | 19 | 19 | 19 |
| Manitoba | 18 | 18 | 19 |
| New Brunswick | 19 | 19 | 19 |
| Newfoundland and Labrador | 19 | 19 | 19 |
| Northwest Territories | 19 | 19 | 19 |
| Nova Scotia | 19 | 19 | 19 |
| Nunavut | 19 | 19 | 19 |
| Ontario | 19 | 19 | 19 |
| Prince Edward Island | 19 | 19 | 19 |
| Quebec | 18 | 18 | 21 |
| Saskatchewan | 19 | 18 | 19 |
| Yukon | 19 | 19 | 19 |

## III. DYNAMICS AND DEPENDENCIES: DEPLOYING AGE VERIFICATION TECHNOLOGIES IN CANADA

While regulations prevail and social needs exist in Canada to enforce age verification through sophisticated technologies, certain factors need to be addressed in this regard, as discussed below:

*A. Accessibility and Inclusiveness*

For any digital technology, accessibility and inclusiveness for all user segments are essential criteria. This is more crucial for age-verification technologies, which address privacy-sensitive information of users. For example, demographic profiles of the user are a significant factor for facial image-based age estimation. The 2021 census reported over 450 ethnic and cultural origins in Canada, with 22.3% of Canadians identified as people of color. So, biometric-based age-verification schemes need to be bias-free for people of different profiles, particularly as racial bias in AI is now a well-known phenomenon [21].

Enhancing trust and confidence among general users for these technologies also needs continuous effort. For instance, onboarding senior citizens to try biometric-based age-verification technologies may first require sufficient awareness building. On the other side of the spectrum, not all parents may be willing to let their children try these solutions. Recent research conducted by the UK Information Commissioner's Office (ICO) and Ofcom present some interesting picture regarding British families' attitude towards age assurance. Most parents there felt that services should have age assurance measures, but these could contradict their desire for control and flexibility over what their children do online. To some parents, age restrictions do not always feel meaningful. So, they were allowing, and even facilitating, their children to circumvent current age assurance measures. Such a scenario re-emphasizes the need to ensure proper acceptance in mass level for age-verification measures [22].

*B. Adaptibility and Transparency*

As Table IV shows, service-specific minimum age requirements can be different for different Canadian provinces. So, any tech solution needs to address this variation if it aims for a Canada-wide deployment. In fact, local context needs to be considered not only by legal requirements, but also to address the usage behaviors. 9.4% of the Canadian population are indigenous people. 17.8% of Canadians live in rural areas. The digital divide in Canada, which is a well-established fact, has caused slower adoption of digital technologies for these communities. This can be more evident for sensitive services like age verification. So, adaptability to local context is a key determinant for any age-verification scheme to be acceptable to the general users [21].

Another important aspect is the minimum age limit. The first law to set a digital age of consent was the Children's Online Privacy Protection Act (COPPA) implemented in the US in 1998, which set the minimum age at 13 years. Originally conceived as a marketing code at a time when the digital world was neither as pervasive nor persuasive as it is now, COPPA sought to restrict advertisers to accessing children under 13 [23]. Since then, 13 years has remained as a standard limit, as used by most social media platforms till now. As a result, children aged 13 to 17 receive no specific protection, creating a de facto age of adulthood online at 13. Recently, in Europe, the GDPR has set the digital age of consent at 16. With rapidly changing context in all aspects of life, this minimum age limit needs to be reviewed regularly in accordance with country and region-specific needs. Overall, the guidelines need to be transparent enough to boost the confidence of both the users and solution developers.

*C. Agility and Accountability*

For researchers and innovators in any country, having favorable regulations and encouraging policies are always important. Canada has been assumed to deal with an "innovation problem" for decades, being ranked 15[th] in the 2022 Global Innovation Index (GII) [24]. To strengthen the innovation ecosystem in Canada, various endeavors have been taken by the government, academic, and private sectors. However, for innovators working on a novel product like age verification, challenges can be multi-fold, as they need to address technical-social-regulatory needs simultaneously. Some agility in regulations can be helpful for them, which will be discussed in detail in the consequential section.

Accountability is another key consideration for the age verification ecosystem. Ensuring regulation for big tech giants, who dominate the digital spaces globally, is now a broadly discussed topic. For online age verification, there always remains a trade-off between state-driven regulation (like 18 years for adult content) and platform-specific regulation (like 13 years for social media). Hence, ensuring accountability can be challenging in scenarios where both regulations address the same issues. This can be mitigated through broader collaboration between governments and private entities.

IV. ONLINE AGE-VERIFICATION IN CANADA: AVAILABLE TECHNICAL STANDARDS

Any technological solution needs to be evaluated and accredited in terms of available standards before making it technically and commercially acceptable. For age-verification technologies, this involves certain dimensions. Firstly, age-verification regulations (e.g., minimum age requirements) vary within different countries-regions and platforms. While there remain some similarities among these, there is still no single global standard. Secondly, this is a broad ecosystem where standards can be generated from global standardization forums (like IEEE, ISO), local regulatory authorities (like UK Information Commissioner's Office), or from certain organizations skilled in this domain (like ACCS, Yoti). Thirdly, different methods of age verification have different attributes, hence they may need different sets of standards. For instance, document-based verification has a simplified procedure than biometric-based verification. Or, within biometric options, facial detection-based schemes require different standards than voice or ECG-based schemes.

Considering the Canadian context, we shall discuss three key standards that could be relevant to researchers and service providers here:

*A. PAS 1296:2018*

PAS 1296:2018 is a standard for online age verification service providers developed by the British Standards Institute and the Digital Policy Alliance. To date, it is considered as the most comprehensive standard in this domain, addressing issues related to privacy, security, safety, usability, accessibility, and data protection. It is applicable to both in-house and third-party age check services for all possible use cases.

. Based on this standard, independent accreditation organizations like ACCS (Age Check Certification Scheme) have developed a certain set of technical requirements for age verification solutions. These standards and processes have been implemented widely in the UK, making them a pioneering hub in this domain. There are currently 56 types of products spanning across 16 sectors, that are age-restricted in the UK [25].

*B. IEEE STD 2089-2021*

IEEE STD 2089-2021 is a standard published by the IEEE Standard Association on 2021, named as the" IEEE Standard for an Age-Appropriate Digital Services Framework Based on the 5Rights Principles for Children". This standard provides a set of processes for digital services when end users are children and aids in the tailoring of the services that are provided so that they are age-appropriate. It also provides a specific impact rating system and evaluation criteria and explains how vendors, public institutions, and the educational sector can meet the criteria. Following this, IEEE has also established an "Online Age Verification Working

Group" to establish a framework for the design, specification, evaluation, and deployment of age verification systems [26].

*C. Pan-Canadian Trust Framework (PCTF)*

The Pan-Canadian Trust Framework (PCTF) is a publicly available set of tools, shared principles, and guidelines as required for digital identity-based services in Canada. It has been developed by the Digital ID & Authentication Council of Canada (DIACC), a consortium of leading entities in this arena, along with public consultation. This is an extensive framework addressing authentication, verification, privacy, credentials, and infrastructures - both technologically and operationally. As a technology-agnostic framework, PCTF is considered to complement existing standards and policies to accelerate the enhancement and adoption of digital ID-based services in Canada [27].

TABLE V. COMPARING AVAILABLE TECHNICAL STANDARDS FOR AGE VERIFICATION IN CANADA

| Criteria | PAS 1296: 2018 | IEEE STD 2089-2021 | PCTF |
|---|---|---|---|
| Flexibility of update | High | Low | Medium |
| Adaptability with local context | Medium | Low | High |
| Depth on specific technology | Medium | High | Low |

As shown in the above table, each of the mentioned standards has a certain relevance for researchers and service providers. Being developed through a long-term process, IEEE STD 2089-2021 has lower flexibility of frequent updates or customization with local context. Being a Canadian standard, PCTF can address local contexts more adequately. For PAS 1296:2018, flexibility can be greater considering its broad exposure to a variety of solutions. However, if we consider which standards may be able to address specific needs for a particular engineering solution, then IEEE STD 2089-2021 and PAS 1296:2018 are better capable. The PCTF will not be a viable option in that regard, as it is a more generalized framework.

In summary, these standards actually complement each other rather than substituting. Let us consider an example. For the ECG-based age verification solution, the below entities or guidelines will be relevant:
* Technical compatibility: IEEE STD 2089-2021
* Performance accuracy: PAS 1296:2018
* Interoperability: PCTF

## V. CONCLUSION

As more countries and regions enforce online age verification, its market size will keep booming. Only in the OECD countries, annual revenues from the age-verification market were estimated as 10.45 billion USD in 2021[1]. Canada is also expected to experience continuous growth in this market, which will facilitate greater investment and further innovation. Regulations and policies will also play a pivotal role here. Some key questions will remain to be addressed on this journey. Like: should age-verification systems be solely regulated by governments, co-regulated or self-regulated by the private sector? A regulatory framework that offers certainty to businesses and builds trust among users will drive the innovations better. Also, age-verification tools should not be mistaken for a silver bullet or a shortcut to making the digital world suitable for children. It is just a single aspect of a broader arena, which can be beneficial for children only through proper support from social and educational endeavors. Basic requirements for any age-verification solution (such as privacy, convenience, security, accessibility, transparency, inclusiveness, rights-respect, and accountability) will not fluctuate. It is likely that as government regulations get enforced, academic researchers develop new technologies, and private entities continue to invest, a more mature ecosystem will emerge for age verification technologies in Canada.

REFERENCES

[1] Age Verification Providers Association, "AV around the world," AVPA [online]. Available: https://avpassociation.com/map/
[2] "But how do they know it is a child? Age Assurance in the Digital World," 5Rights Foundation, Oct. 2021. [Online]. Available: https://5rightsfoundation.com/in-action/but-how-do-they-know-it-is-a-child-age-assurance-in-the-digital-world.htm
[3] "Digital Economy Act 2017," *Legislation.gov.uk*, 2017. [Online]. Available: https://www.legislation.gov.uk/ukpga/2017/30/contents/enacted
[4] UK Department for Digital, Culture, Media & Sport, "Online Safety Bill", *UK Parliamentary Bills*, Jan. 18, 2023. [Online]. Available: https://bills.parliament.uk/bills/3137
[5] "Public Bill (Senate) S-203 (43-2) - - Protecting Young Persons from Exposure to Pornography Act - Parliament of Canada", *Parl.ca*, 2022 [Online]. Available: https://www.parl.ca/DocumentViewer/en/43-2/bill/S-203/third-reading
[6] "No porn, no Instagram for kids: France doubles down on age verification," *POLITICO*, Feb. 15, 2023. [Online]. Available: https://www.politico.eu/article/no-porn-no-instagram-for-kids-france-doubles-down-age-verification-emmanuel-macrons-nick-clegg/.
[7] L. Pasquale, P. Zippo, C. Curley, B. O'Neill and M. Mongiello, "Digital Age of Consent and Age Verification: Can They Protect Children?" in *IEEE Software,* vol. 39, no. 3, pp. 50-57, May-June 2022.
[8] Yoti, "Yoti Age Estimation White Paper Full version," 2022 [Online]. Available: https://www.yoti.com/wp-content/uploads/Yoti-Age-Estimation-White-Paper-May-2022.pdf
[9] "Instagram to test 'facial age estimation technology' in Canada," *CTVNews*, Mar. 02, 2023. [Online]. Available: https://www.ctvnews.ca/sci-tech/instagram-to-test-facial-age-estimation-technology-in-canada-1.6295327
[10] S. Safavi, M. Russell, and P. Jancovic, "Automatic Speaker, Age-group and Gender Identification from Children's Speech," *Computer Speech & Language*, vol. 50, pp. 141–156, 2018.
[11] A. S. Falohun, O. D. Fenwa, and F. A. Ajala, "A Fingerprint-based Age and Gender Detector System using Fingerprint Pattern Analysis," *Int. J. of Computer Applications*, vol. 136, no. 4, 2016.
[12] D. Yaman, F. I. Eyiokur, N. Sezgin, and H. K. Ekenel, "Age and Gender Classification from Ear Images." *Proc. 2018 Int. Workshop on Biometrics and Forensics* , pp. 1-7.
[13] M. Erbilek, M. Fairhurst, and M. C. D. C. Abreu, "Age Prediction from Iris Biometrics." *Proc. 2013 5th Int. Conf. on Imaging for Crime Detection and Prevention*, pp. 1-5.
[14] A. Adib, W. -P. Zhu and M. O. Ahmad, "Adult and Non-Adult Classification Using ECG," 2022 IEEE 7th Forum on Research and Technologies for Society and Industry Innovation (RTSI), Paris, France, 2022, pp. 174-179
[15] S. C. Government of Canada, "The Daily — Access to the Internet in Canada, 2020," *www150.statcan.gc.ca*, May 31, 2021. [Online]. Available: https://www150.statcan.gc.ca/n1/daily-quotidien/210531/dq210531d-eng.htm

[16] "Social Media Statistics in Canada for 2023 - Made in CA," Feb. 23, 2023. [Online]. Available: https://madeinca.ca/social-media-statistics-canada/

[17] "Alcohol & Cannabis," *Retail Council of Canada*, Mar. 14, 2023. [Online]. Available: https://www.retailcouncil.org/alcohol-cannabis/.

[18] "Gambling Statistics in Canada for 2023 - Made in CA," Aug. 07, 2022. [Online]. Available: https://madeinca.ca/gambling-canada-statistics/

[19] "Pornhub releases Canada's top searches of the past year | News," *dailyhive.com*. [Online]. Available: https://dailyhive.com/vancouver/pornhub-canada-year-review.

[20] "Under 19 and traveling to Canada? Age of majority and legal age 101," *pvtistes.net*, Aug. 19, 2022. [Online]. Available: https://pvtistes.net/en/canada-age-of-majority-legal-age/.

[21] Statistics Canada, "The Daily — The Canadian census: A rich portrait of the country's religious and ethnocultural diversity," *www150.statcan.gc.ca*, Oct. 26, 2022. [Online]. Available: https://www150.statcan.gc.ca/n1/daily-quotidien/221026/dq221026b-eng.

[22] Information Commissioner's Office (ICO) UK and Ofcom, "Family's attitudes towards age assurance," Oct. 2022 [Online]. Available: https://www.gov.uk/government/publications/families-attitudes-towards-age-assurance-research-commissioned-by-the-ico-and-ofcom

[23] Federal Trade Commission, "Children's Online Privacy Protection Rule ('COPPA')," *Federal Trade Commission*, Jul. 25, 2013. [Online]. Available: https://www.ftc.gov/legal-library/browse/rules/childrens-online-privacy-protection-rule-coppa

[24] WIPO, "Global Innovation Index 2022, 15th Edition," *WIPO*, 2022. [Online]. Available:https://www.wipo.int/global_innovation_index/en/2022

[25] "PAS 1296 - Age Check Certification Scheme," *www.accscheme.com*, Mar. 31, 2018. [Online]. Available: https://www.accscheme.com/services/age-assurance/pas-1296.

[26] "IEEE SA - IEEE 2089-2021," *IEEE Standards Association*, Nov. 09, 2021. [Online]. Available: https://standards.ieee.org/ieee/2089/7633/

[27] Digital ID & Authentication Council of Canada, "PCTF Overview," *diacc.ca*, Oct. 2022. [Online]. Available: https://diacc.ca/trust-framework/pctf-overview/